\documentclass[aps, prd, twocolumn, amsmath, amssymb, floats, floatfix, groupedaddress, nofootinbib, nopacs,notitlepage]{revtex4}

\usepackage{latexsym}
\usepackage{graphicx}
\usepackage{amsmath}
\usepackage{amssymb}
\usepackage{amsfonts}
\usepackage{times}
\usepackage{subfigure}		
\usepackage[normalem]{ulem} 

\usepackage{graphicx}
\usepackage{dcolumn}
\usepackage{bm}
\usepackage{tikz}
\usepackage{bigints}
\usepackage{array,tabularx,multirow,booktabs}
\usepackage{tabularx}
\usepackage{multirow}
\usepackage{amssymb,mathtools,pgfplots,pgfmath}

\newcolumntype{z}{>{\centering\arraybackslash}X}

\DeclareMathOperator{\arcsecc}{arcsec}
\DeclareMathOperator{\diag}{diag}

\usepackage[tracking=true]{microtype}
\SetTracking{}{500}
\SetTracking{encoding={*}, shape=sc}{40}

\begin{document}

\title{Weak gravitational lensing by two-power-law densities using the Gauss-Bonnet theorem}



\author{Karlo de Leon}
\email{kndeleon@nip.upd.edu.ph}
%
\author{Ian Vega}
\email{ivega@nip.upd.edu.ph}
\affiliation{
 National Institute of Physics,
 University of the Philippines, Diliman, Quezon City, 1101, Philippines
 }

\begin{abstract}
We study the weak-field deflection of light by mass distributions described by two-power-law densities $\rho(R)=\rho_0 R^{-\alpha}(R+1)^{\beta-\alpha}$, where $\alpha$ and $\beta$ are non-negative integers. New analytic expressions of deflection angles are obtained via the application of the Gauss-Bonnet theorem to a chosen surface on the optical manifold. Some of the well-known models of this two-power law form are the Navarro-Frenk-White (NFW) model $(\alpha,\beta)=(1,3)$, Hernquist  $(1,4)$, Jaffe $(2,4)$, and the singular isothermal sphere $(2,2)$. The calculated deflection angles for Hernquist and NFW agrees with that of Keeton and Bartelmann, respectively. The limiting values of these deflection angles (at zero or infinite impact parameter) are either vanishing or similar to the deflection due to a singular isothermal sphere. We show that these behaviors can be attributed to the topological properties of the optical manifold, thus extending the pioneering insight of Werner and Gibbons to a broader class of mass densities.
\end{abstract}

\pacs{Valid PACS appear here}
\maketitle


\section{Introduction}
Gravitational lensing has come a long way since its entry to modern science. Getting its spotlight with the solar eclipse of 1919, Eddington's expedition to firm up Einstein's general relativity on more solid empirical ground (notwithstanding the ensuing historical controversy \citep{coles2001einstein,kennefick2012not}) is now the stuff of scientific folklore. This bond between lensing and general relativity has only grown stronger in the years since, aided by increasingly more sophisticated instruments and techniques \citep{lebach1995measurement,reyes2010confirmation}. Meanwhile, gravitational lensing has far outgrown its status as a mere theoretical prediction. It is now an indispensable tool for much of modern astrophysics and cosmology, serving as a primary probe for characterizing mass distributions throughout the cosmos \citep{tyson1990detection,metcalf2001compound,PhysRevD.98.103517,birrer2019h0licow,10.1093/mnras/stw2805,abbott2018dark,diehl2014dark,hu2002mass}, particularly in the high redshift regime \citep{kelly2018extreme,fian2018estimate,salmon2018relics,acebron2018relics,lamarche2018resolving,zavala2018dusty}.

This paper returns to lensing's classic roots, by focusing on the relation between the lensing behavior of a galaxy and its mass distribution. A lens model is an important initial assumption in inverting lensed images back to its source image \citep{kayser1988imaging,warren2003semilinear}. In astrophysics, knowing the expected lensing behavior of a density model is essential in testing its applicability for modelling mass clusters. Generally, the lensing properties of a density function, such as the deflection angle and magnification, are not readily solvable. Mass models are then typically chosen based on how readily observables can be calculated from them. 

Many of the commonly used density functions for galaxies and dark matter halos belong to the family of density parametrizations whose generalized form first appeared in a paper by Hernquist \citep{hernquist1990analytical}:
\begin{equation}\label{eq:rho_hernquist_general_af}
    \rho(R) = \frac{\rho_0}{R^{\alpha}(R^{1/\gamma}+1)^{(\beta-\alpha)\gamma}}
\end{equation}
where $R=r/r_0$. This has become a common choice for modelling due to its relatively simple form and its analytic properties \citep{dehnen1993family,tremaine1994family,zhao1996analytical}. Here, there are two scale parameters $r_0$ and $\rho_0$, and three exponential parameters $(\alpha,\beta,\gamma)$ that modify the general shape of the distribution. Central growth is controlled by $\alpha$: $\rho\sim R^{-\alpha}$ for small $R$. This takes into account the central cusp observed on the surface brightness profiles of some galaxies, even at high resolution imaging \citep{lauer1992planetary,lauer1993planetary,faber1997centers}. The allowed divergence is restricted to values $\alpha<3$, so that the mass function \eqref{eq:mass_function} may still be defined.  Meanwhile, radial decay is regulated by $\beta$: $\rho\sim R^{-\beta}$ for large $R$. The density profile \eqref{eq:rho_hernquist_general_af} is a function that provides a smooth transition between these two power-laws, with the exponent $\gamma$ measuring the width of the transition region.

Here, we study the weak gravitational lensing of the so-called two-power-law densities, a subset of the Hernquist family:
\begin{equation}\label{eq:rho_general}
\rho(R) = \frac{\rho_0}{R^{\alpha}(R+1)^{\beta-\alpha}}
\end{equation}
In particular, we calculate the deflection angle in the high-frequency and weak-field limit, with the source and observer both at spatial infinity. It turns out that for integer values of $\alpha$ and $\beta$, the deflection angle can be expressed analytically, and so we limit our discussion to these values. Models belonging to this set include the famous Navarro-Frenk-White (NFW) model $(\alpha,\beta)=(1,3)$, Hernquist $(1,4)$, Jaffe $(2,4)$, and the singular isothermal sphere $(2,2)$ \citep{Navarro1993,hernquist1990analytical,jaffe1983simple}. Previous calculations have worked out deflection angles arising from densities related to this form, but with significant restrictions, such as on the $(\alpha,\beta,1/2)$-subset that closely resembles the two-power law \cite{chae1998gravitational,chae2002fast}, a much restricted range of the two-power law \citep{keeton2001catalog}, and on individual values of $\alpha$ and $\beta$ \citep{bartelmann1996arcs}.

Our calculation utilizes the Gauss-Bonnet method by Gibbons and Werner \cite{Gibbons2008}, which nicely highlights the often overlooked role of topology in gravitational lensing. This seminal work motivated us to understand the extent to which the topological arguments made by \cite{Gibbons2008} generalize to a much broader class of density functions. Our results shall show that, indeed, gross features of weak lensing are due to topological properties of an underlying \textit{optical manifold}. Beyond this question of principle, we also argue that for weak lensing of (at least) spherically symmetric distributions, the Gauss-Bonnet approach holds a number of advantages over common methods such as the thin-lens approximation and direct calculations based on metric components. (We shall say more about this in Section \ref{sec:surface}.) Previous works have exploited these advantages to study weak lensing in various contexts \cite{jusufi2017deflection,jusufi2018gravitational,crisnejo2018weak}. Though curiously, almost none of the extant literature applies to model density functions that are particularly useful for astrophysical work. Our work partly seeks to rectify this state of affairs. 

The rest of this paper shall proceed as follows: first, we go over some preliminaries, particularly the Gauss-Bonnet theorem based on the optical metric, and a short summary of Gibbons' and Werner's method. New expressions for the deflection angles of the densities are then derived, and this is followed by a discussion of general observations and comparisons. This brings to the fore the perspective advocated by Gibbons and Werner that gross physical features of weak lensing are primarily determined by the topology and geometry of the underlying optical manifold. The examples we explicitly work out all lend further credence to this point of view. Finally, the paper concludes with a summary and recommendations for future work. 

We will use the signature $(-,+,+,+)$, and geometric units wherein $c=G=1$ throughout.

\section{Gauss-Bonnet Theorem}
In the Gibbons-Werner approach to weak lensing, the deflection angle is directly calculated from the well-known Gauss-Bonnet theorem of classical differential geometry. The theorem has made many appearances in various fields of physics. (See for example \citep{doi:10.1080/00268979500102241,Ryder_1991,yang2015geometric,banados1994black,van2018einstein}, just to name a few.) For completeness, we briefly review this here.

Let $M$ be a compact, oriented (and thus, triangularizable) surface with a piecewise-smooth boundary $\partial M$, where the curve is arclength-parametrized and traversed in the positive sense. The Gauss-Bonnet theorem then states that
\begin{equation}\label{eq:gaussbonnet}
\iint_{M} K \,dS \space + \int_{\partial M} \kappa \,dt + \sum_{i} \alpha_{i} = 2\pi \chi(M)
\end{equation}
where $t$ is the arclength parameter, $\alpha_i$ are the external angles at the vertices of $\partial M$, $K$ is the Gaussian curvature, $\kappa$ is the geodesic curvature, and $\chi$ is the Euler characteristic of $M$ (e.g., \cite{klingenberg2013course}, p. 139). The geodesic curvature of a smooth curve $\gamma$, with unit tangent vector $\dot{\gamma}$ and unit acceleration vector $\ddot{\gamma}$, is defined as
\begin{equation}\label{eq:kappa}
\kappa = g(\nabla_{\dot{\gamma}}\dot{\gamma},\ddot{\gamma})
\end{equation}
while the Gaussian curvature is proportional to the single non-trivial component of the Riemann curvature tensor for two dimensions (Gauss's Theorema Egregium):
\begin{equation}\label{eq:gaussian_curvature}
    K=R_{1212}/|g|
\end{equation}
with $|g|$ the determinant of the metric \citep[pp. 64, 78]{klingenberg2013course}.

The theorem is applied to a choice of surface $D$ defined on the optical metric space, which then generates an expression involving the deflection angle.

\section{Optical metric and the weak-field limit}
The metric of a static and spherically symmetric spacetime has the general form in polar charts:
\begin{equation}\label{eq:metric_spacetime}
ds^2 = -e^{2A(r)}dt^2 + e^{2B(r)}dr^2 + r^2(d\theta ^2 + \sin\theta d\phi ^2) 
\end{equation}
Spherical symmetry guarantees that geodesics lie on a plane and equivalent up to spatial rotations about the origin. Thus, we can set any geodesic to lie on the equatorial plane $\theta=\pi/2$, without loss of generality. Working only with null paths allows us to further reduce the number of coordinates by considering another manifold with the spacelike coordinate $t$ defined as the new interval. From the metric $g_{\mu\nu}$ \eqref{eq:metric_spacetime}, we consider the conformal transformation $\Tilde{g}_{\mu\nu}=g_{\mu\nu}/g_{00}$. With this metric, we set $ds^2=0$, and define $t$ as the new interval.
\begin{equation}\label{eq:metric_opt}
dt^2 = \Tilde{g}_{ab} \, dx^a dx^b = e^{2(B(r)-A(r))}dr^2 + e^{-2A(r)}r^2d\phi ^2
\end{equation}
This is the optical metric $g_{\textrm{opt}}=g_{ab}/g_{00}$. While geodesics in general are not preserved under conformal transformations, it does hold for null curves (e.g., \cite{wald2010general}, p. 446). So, light paths are still faithfully represented by geodesic curves.

For the perfect fluid case, the energy-momentum tensor is $T_{\mu\nu}=\diag{(\rho,p,p,p)}$ in the rest frame of the fluid, where $\rho$ is the energy density and $p$ is the isotropic rest-frame pressure. Plugging this to Einstein's field equations, the metric components are computed from the energy-momentum tensor components as
\begin{equation}\label{eq:A}
\frac{dA}{dr} = \bigg( 1 - \frac{2m(r)}{r} \bigg) ^{-1} \bigg(\frac{m(r)}{r^2} + 4\pi Grp \bigg)
\end{equation}
\begin{equation}\label{eq:B}
e^{-2B(r)} = 1 - \frac{2m(r)}{r}
\end{equation}
where $m(r)$ is the mass function
\begin{equation}\label{eq:mass_function}
    m(r) := \int_{0}^{r} \rho(r') \, 4\pi {r'}^2 \, dr'
\end{equation}
(e.g., \cite{schutz2009first}, pp. 261-262). With the specification of an equation of state or, in our case, a density function, the pressure can be calculated from the Tolman-Oppenheimer-Volkoff (TOV) equation
\begin{equation}\label{eq:tov}
    \frac{dp}{dr} = - \frac{(\rho+p)(m+4\pi r^3p)}{r^2 (1-2(m/r))}
\end{equation}
derived from the Einstein field equations and the conservation of energy-momentum tensor, which is also a consequence of the former \citep[p. 264]{schutz2009first}. These equations, plus boundary conditions, completely define the metric due to the perfect fluid. 

In terms of the physical parameters of the density function \eqref{eq:rho_general}, the weak-field limit is defined as the low-density case $\mu:=\rho_0 r_0^2 \ll 1$ keeping terms only up to first order in $\mu$. Pressure contributions may be neglected in this limit. Expanding the pressure term $pr_0^2$ in equation \eqref{eq:tov} in powers of $\mu$ (note that $\rho r_0^2\propto\mu$ and $m/r_0\propto \mu$), we find that it is constant at first order. The pressure is expected to vanish at spatial infinity, and so it must also vanish everywhere. Meanwhile, the zeroth order is zero because $p$ must vanish for a vacuum. Note however that we can only do the expansion in equation \eqref{eq:tov} and approximate $p=0$ everywhere if $2m(r)/r\ll1$ for all $r\geq0$. This holds for most $\alpha,\beta$ given a sufficiently small value of $\mu$. When $\beta<2$, the radial gradient of the pressure blows up for some finite radius $r_{\textrm{crit}}$, which signifies an infinite amount of force exerted on the infinitesimal spherical shell at $r=r_{\textrm{crit}}$. We will not consider this case here.

\section{Surface construction}\label{sec:surface}
Here we give a short review of the surface designed by Gibbons and Werner \cite{Gibbons2008} for calculating deflection angles through the Gauss-Bonnet theorem. Note however that such surface constructions are not unique (e.g., \cite{ishihara2016gravitational,arakida2018light}).

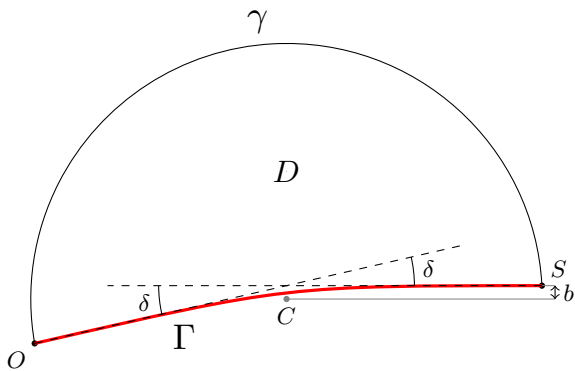
\begin{figure}[h!]
\centering
\begin{tikzpicture}[scale=3.4]
\path[name=circ] (1,0) arc[start angle=0, end angle=190, radius=1];
\draw[gray, very thin] (0,0) -- (1.05,0);
\draw (3:1) arc [start angle=3, end angle=190, radius=1] node[midway, above, scale=1.5]{$\gamma$};
\filldraw[gray] (0,0) circle[radius=0.01] node[below, black, scale=1]{$C$};
\filldraw[black] (3:1) circle[radius=0.01] node[anchor=south west, scale=1]{$S$};
\filldraw[black] (190:1) circle[radius=0.01] node[anchor=north east, scale=1]{$O$};
\draw[red,very thick] (3:1) .. controls (0,{sin(3)}) .. node[black, below, near end, scale=1.5]{$\Gamma$} (190:1);
\draw[dashed] (3:1) -- (-0.7,{sin(3)});
\draw[dashed] (190:1) -- (0.7,{sin(3)-0.7*((sin(3)-sin(190))/(cos(190)))});

\draw (0.5, {sin(3)}) node[anchor=south west, scale=1]{$\delta$} arc[start angle=0, end angle=12.92393662, radius=0.5];
\draw (-0.5, {sin(3)}) node[anchor=north east, scale=1]{$\delta$} arc[start angle=180, end angle=192.92393662, radius=0.5];

\draw[gray, very thin] (3:1) -- +(0.05,0);
\draw[<->] (1.05,0) -- node[anchor=west]{$b$} +(0,{sin(3)});

\draw (0,0.499) -- (0,0.5) node[scale=1.3]{$D$};
\end{tikzpicture}
\caption{Surface on the optical manifold used for computing the deflection angle.}
\label{fig:surface}
\end{figure}

On the chart map of the optical manifold, let the origin of the coordinate system $(r,\phi)$ be at the center $C$ of the mass distribution (see Figure \ref{fig:surface}). The geodesic curve $\Gamma$ is the trajectory of the photon emitted at the source $S$ with impact parameter $b$ received by an observer at $O$. Further, let $S$ and $O$ be at an equal distance $d$ from $C$. The angle $\delta$ between the tangent of $\Gamma$ at $O$ and the line $\phi=\pi$ is the deflection angle $\delta$. We construct a surface from $\Gamma$ by considering an additional circular arc $\gamma$ centered at $C$ intersecting $\Gamma$ at points $S$ and $O$. Let $D$ be this surface bounded by $\Gamma$ and $\gamma$. Finally, we take the limit $d\rightarrow\infty$ so that $S$ and $O$ are at spatial infinity, and $\phi(S)=0$ and $\phi(O)=\pi+\delta$.

The Gauss-Bonnet theorem on surface $D$ reads
\begin{multline}\label{eq:master_exact}
\int_{0}^{\pi+\delta}\int_{(r\circ\Gamma)(\phi)}^{\infty} K \sqrt{|g_{\textrm{opt}}|} \, \,dr \, d\phi \, \\+ \int_{0}^{\pi + \delta} \bigg(\kappa \frac{dt}{d\phi} \bigg) \, d\phi \, + \bigg( \frac{\pi}{2}+\frac{\pi}{2}\bigg)=  2\pi
\end{multline}
noting that the differential element $dS$ in coordinate form is
\begin{equation}\label{eq:dS}
    dS = \sqrt{|g_{\textrm{opt}}|}\,dr\,d\phi
\end{equation}
and the external angles at $S$ and $O$ are $\pi/2$. The surface $D$ does not contain the possibly singular point $C$, so the surface is simply connected and has an Euler characteristic $\chi=1$. The Gaussian curvature can be solved from equation \eqref{eq:gaussian_curvature} given the metric $g_{\textrm{opt}}$: 
\begin{multline}\label{eq:K_exact}
-K \sqrt{|g_{\textrm{opt}}|} = \frac{2m}{r^2} \Big( 1 - \frac{2m}{r} \Big) ^{-3/2} \times \\ \Bigg( 1 - \frac{3m}{2r} - \frac{4\pi\rho r^3}{m} \Big( 1 - \frac{2m}{r} \Big) \Bigg) 
\end{multline}
In the weak-field limit, equation \eqref{eq:K_exact} is solved only up to first order in $\mu$. It then suffices to take only the zeroth order of the geodesic curve $(r\circ\Gamma)(\phi)=b/\sin{\phi}$ (the undeflected light curve in Minkowski space), and the zeroth order of the angular bound: $\pi+\delta \approx \pi$ (we know that $\delta=0$ in vacuum, so $\delta$ must be at least of order $\mathcal{O}(\mu)$). With this, we obtain the central equation for calculating deflection angles
\begin{equation}\label{eq:master}
\int_{0}^{\pi + \delta} \bigg(\kappa \frac{dt}{d\phi} \bigg) \, d\phi - \pi = \int_{0}^{\pi}\int_{B/\sin \phi}^{\infty} \mathcal{K} \, \,dR \, d\phi 
\end{equation}
where $B=b/r_0$, and $\mathcal{K}$ is the zeroth order of equation \eqref{eq:K_exact} with a multiplicative constant $r_0$
\begin{equation}\label{eq:integrand}
\mathcal{K} :=  \frac{2}{R^2} \frac{m}{r_0} - 8\pi R r_{0}^2 \rho
\end{equation}
It is this form $\mathcal{K}$, rather than $K$, that is directly useful for our calculations. We will call $\mathcal{K}$ the Gaussian curvature \textit{term}.

This method offers a number of advantages in calculating deflection angles from spherical matter distributions compared to canonical methods. Integration from $g(p^{\mu},p^{\mu})=0$ (e.g., \cite{schutz2009first}, pp. 283-284), where $p^{\mu}$ is the four-momentum of the photon, requires the analytic form of the spacetime metric components. While $g_{rr}$ is readily obtained from the mass function, $g_{tt}$ will have to be computed from the TOV equation \eqref{eq:tov}, where analytic form may not be guaranteed. The method of thin-lens approximation partially resolves this problem, since it only requires the energy density function (e.g., \cite{mollerach2002gravitational}, p. 25). The difficulty however is translated to computing the surface mass density, which is the projection of the mass distribution on a plane orthogonal to the light ray direction $\Sigma = \int\rho\,\mathrm{d}z$. Because the spherical energy density is a function of $r=\sqrt{x^2+y^2+z^2}$, the integrand $\rho\circ r$ can easily have a complicated form in $z$, even for fairly simple functions $\rho(r)$. In the Gauss-Bonnet method, the central equation, \eqref{eq:master} and \eqref{eq:integrand}, only requires the explicit forms of the density and mass function, and the integration is already carried in the $r$ and $\theta$ space. For some spacetimes, $g_{tt}$ must still be calculated from the TOV equation \eqref{eq:tov}. But, only the $r\rightarrow\infty$ limit is needed, so the TOV equation further simplifies.

Aside from advantages in calculation, the method also gives insight to the role of the topology in gravitational lensing, as we will see later.

\section{Calculation of deflection angles}

A general expression of the deflection angle that covers all values of $\alpha,\beta$ was not obtained due to the non-homotopy of the function $f(x,t) = \int^x y^{a-t(a+1)} \,dy$. The calculation splits into separate cases whenever a $\int^x y^{-1} \,dy$-integration occurs. We present the deflection angles in decreasing order of $\alpha$, and then in decreasing values of $\beta$. Most of the calculations are similar and involve only the same family of integrals. The only significant difference is between the cases $\beta>2$ and $\beta=2$. Densities with $\beta>2$ are asymptotically flat, while $\beta=2$ approaches the singular isothermal distribution at infinity. Before proceeding, we present first the three integrals that repeatedly appear in our calculations:
\begin{widetext}
\begin{align}
I_1^q&:=\int_{0}^{\pi} \frac{\sin{\phi}\,\,d\phi}{(1+B\csc \phi)^{q}} = \frac{1}{B^{q}} \frac{(-1)^{q-1}}{(q-1)!} \bigg( \frac{\partial}{\partial a}  \bigg) _{a=1/B}^{q-1} \Bigg( \frac{2}{a} - \frac{\pi}{a^2}\bigg( 1-\frac{1}{\sqrt{1-a^2}} \bigg) - \frac{2\arcsin{a}}{a^2\sqrt{1-a^2}} \Bigg)\label{eq:I2}\\
I_2^q&:=\int_{0}^{\pi} \frac{d\phi}{(1+B\csc \phi)^{q}} = \frac{1}{B^{q}} \frac{(-1)^{q-1}}{(q-1)!} \bigg( \frac{\partial}{\partial a}  \bigg) _{a=1/B}^{q-1} \Bigg( \frac{\pi}{a}\bigg( 1-\frac{1}{\sqrt{1-a^2}} \bigg) + \frac{2\arcsin{a}}{a\sqrt{1-a^2}} \Bigg)\label{eq:I1}\\
I_3 &:= \int_{0}^{\pi} \sin{\phi} \ln{(1+B\csc{\phi})}\,\,d\phi=\pi B + 2\ln \frac{B}{2} - 2\sqrt{B^2-1}\arctan{\sqrt{B^2-1}}\label{I_3}
\end{align}
\end{widetext}
We will derive these in the appendix.

\subsection{Case $(2,\beta\geq4)$}
For asymptotically flat metric, the deflection angle is solely due to the area integral in equation \eqref{eq:master}:
\begin{equation}\label{eq:deflect_area}
\delta = \int_{0}^{\pi} \int_{B/\sin \phi}^{\infty} \mathcal{K} \,dR \,d\phi\,
\end{equation}
since in flat space the geodesic curvature $\kappa$ of a circular arc is the usual inverse of the radius $\kappa=1/d$, and $dt/d\phi=d$. The energy density and mass function are
\begin{gather}
\rho(R) = \frac{\rho_0}{R^2(R+1)^{\beta-2}}\label{eq:rho_p2q}\\
m(R) = \frac{4 \pi \rho_{0} r_{0}^3}{\beta-3} \bigg( 1 - \frac{1}{(R+1)^{\beta-3}}\bigg)\label{eq:mass_p2q}
\end{gather}
Therefore, the Gaussian curvature term gives
\begin{multline}\label{eq:K_p2q_pre}
\mathcal{K} =  8 \pi \rho_{0} r_{0}^2 \bigg( \frac{1}{\beta-3}\frac{1}{R^2} - \frac{1}{\beta-3} \frac{1}{R^2(R+1)^{\beta-3}} \\- \frac{1}{R(R+1)^{\beta-2}}\bigg)
\end{multline}
But, this form of $\mathcal{K}$ is not fit for the $R$-integration. Our workaround is to perform partial fraction decomposition. In the appendix, we present a general decomposition of the fraction $(R-a)^{-p}(R-b)^{-q}$. The following results are used to decompose fractions in equation \eqref{eq:K_p2q_pre}:
\begin{gather}
\frac{1}{R(R+1)^{j}} = \frac{1}{R} - \sum_{i=1}^{j} \frac{1}{(R+1)^i}\label{eq:decompose_p1}\\
\frac{1}{R^2(R+1)^{j}} = -\frac{j}{R} + \frac{1}{R^2} + \sum_{i=1}^{j} \frac{j+1-i}{(R+1)^i}\label{eq:decompose_p2}
\end{gather}
With these, we proceed with the integration and find the deflection angle to be
\begin{multline}\label{eq:deflect_p2q}
\delta(B) = \frac{8\pi \rho_{0} r_{0}^2}{\beta-3} \bigintssss_0^\pi\Bigg(\sum_{i=1}^{\beta-3} \frac{1}{(1+B\csc{\phi})^i}\Bigg)\,d\phi \\= \frac{8\pi \rho_{0} r_{0}^2}{\beta-3}\bigg(\frac{2}{B}-\frac{I_1^{\beta-3}}{B}\bigg)
\end{multline}
The summation in equation \eqref{eq:deflect_p2q} is just a finite geometric sum. It is apparent from this form that $\delta(0)=8\pi^2\rho_{0} r_{0}^2$ and $\delta(\infty)=0$. This non-vanishing deflection at $B=0$ is discussed in Section \ref{sec:discussion}. For the Jaffe model $(2,2)$, the deflection angle is simply
\begin{equation}\label{eq:deflect_Jaffe}
    \delta_{\textrm{Jaffe}}(B)=8\pi\rho_{0}r_{0}^2\Bigg(\pi-\frac{2B}{\sqrt{B^2-1}}\arcsecc{b}\Bigg)
\end{equation}

Deflection angles for a fairly general subset of $(\alpha,\beta,1)$ densities written in terms of analytic functions, such as equation \eqref{eq:deflect_p2q}, may not exist elsewhere in other literature. In checking the validity of the expressions, we therefore inspect the limiting values at $B\rightarrow0$ and $B\rightarrow\infty$ and analyze their corresponding implications.

\subsection{Case (2,3)}
The mass function of this distribution blows up at infinity, yet the deflection angle is still well-behaved. Densities with $\beta>2$ are asymptotically flat, and the deflection angle is solved again from equation \eqref{eq:deflect_area}. The procedure is similar with the previous case, so we only present the results here:
\begin{equation}\label{eq:rho_p2q1}
    \rho(R) = \frac{\rho_0}{R^2(R+1)}
\end{equation}
\begin{equation}\label{eq:mass_p2q1}
m(R) = 4\pi\rho_{0}r_{0}^3 \ln{(R+1)}
\end{equation}
\begin{equation}\label{eq:deflect_p2q1}
\delta(B) = 8\pi\rho r_{0}^2\frac{I_3}{B}
\end{equation}
Similar to the previous case, $\delta(0)=8\pi^2\rho_{0} r_{0}^2$ and $\delta(\infty)=0$.

\subsection{Case $(1,\beta\geq4)$}
We start from equation \eqref{eq:deflect_area} with the following density and mass function:
\begin{equation}\label{eq:rho_p1q}
\rho(R)=\frac{\rho_0}{R(R+1)^{\beta-1}}
\end{equation}
\vspace{-25pt}
\begin{multline}\label{eq:mass_p1q}
m(R) = 4 \pi \rho_{0} r_{0}^3 \Bigg( \frac{1}{(\beta-2)(\beta-3)} \\- \frac{1}{\beta-3} \frac{1}{(R+1)^{\beta-3}} + \frac{1}{\beta-2} \frac{1}{(R+1)^{\beta-2}} \Bigg)
\end{multline}
The Gaussian curvature term can be written as
\begin{multline}\label{eq:K_p1q}
\mathcal{K} =  8\pi \rho_{0} r_{0}^2 \Bigg(-\frac{1}{(R+1)^{\beta-1}} \\+ \frac{1}{(\beta-3)(\beta-2)}\sum_{i=2}^{\beta-2}\frac{i-1}{(R+1)^i} \Bigg)
\end{multline}
using again the relations \eqref{eq:decompose_p1} and \eqref{eq:decompose_p2}. Proceeding further, the deflection angle is
\begin{equation}\label{eq:deflect_p1q}
\delta(B)=\frac{8\pi \rho_{0} r_{0}^2}{(\beta-2)(\beta-3)} \Bigg(\frac{2}{B}-\frac{I_1^{\beta-2}}{B}-(\beta-2)I_2^{\beta-2}\Bigg)
\end{equation}
where $\delta(0)=\delta(\infty)=0$. Thus, the deflection angle of the Hernquist model $(1,4)$ is
\begin{equation}\label{eq:deflect_Hern}
    \delta_{\textrm{Hern}}(B)=\frac{8\pi\rho_{0}r_{0}^2 B}{B^2-1}\bigg(1-\frac{\arcsecc{b}}{\sqrt{B^2-1}}\bigg)
\end{equation}
This is similar to the result presented in the catalog of Keeton \cite{keeton2001catalog}.

\subsection{Case $(1,3)$ (NFW)}
Similar to the density $(2,3)$, the mass function of NFW diverges at infinity. The deflection angle can still be solved, however. The density and mass function of NFW are given by
\begin{gather}
\rho(R)=\frac{\rho_0}{R(R+1)^2}\label{eq:rho_NFW}\\
m(R) = 4\pi\rho_{0}r_{0}^3 \bigg( \ln{(R+1} - \frac{R}{R+1} \bigg)\label{eq:mass_NFW}
\end{gather}
Solving equation \eqref{eq:deflect_area} yields
\begin{multline}\label{eq:deflect_NFW}
\delta_{\textrm{NFW}}(B)=8\pi\rho_0r_{0}^2\bigg(\frac{I_3}{B}-I_2^{1}\bigg)\\=\frac{16\pi\rho r_{0}^2}{B} \bigg(\ln \frac{B}{2} + \frac{1}{\sqrt{B^2-1}}\arctan \sqrt{B^2-1} \bigg)
\end{multline}
The deflection angle is defined for all $B$ and has the same limits as the previous case: $\delta(0)=\delta(\infty)=0$. This is equivalent to the result of Bartelmann \cite{bartelmann1996arcs}.

\subsection{Case $(1,2)$}

Densities with $\beta=2$ have divergent mass functions at infinity and are not asymptotically flat. Unlike the previous cases where the deflection angle is only due to the area integral, here we have a contribution from the circular arc $\gamma$. We return to equation \eqref{eq:master} dealing first with the area integration, then the line integral part.

Evaluating the area integral proceeds similarly:
\begin{equation}\label{eq:rho_p1q1}
    \rho = \frac{\rho_0}{R(R+1)}
\end{equation}
\begin{equation}\label{eq:mass_p1q1}
    m(R) = 4 \pi \rho_{0} r_{0}^3 \big(R - \ln{(R+1)}\big)
\end{equation}
\begin{equation}\label{eq:area_part_p1q1}
    \delta_{\textrm{area}}(B) = -8\pi\rho r_{0}^2\frac{I_3}{B}
\end{equation}

Now, we compute the metric coefficients to evaluate the line integral. With the given density and mass function, \eqref{eq:rho_p1q1} and \eqref{eq:mass_p1q1}, equations \eqref{eq:A} and \eqref{eq:B} are solved up to first order in $\mu$ giving
\begin{gather}
e^{2A} \approx C^{-2} (R+1)^{8\pi \rho_0 r_0^{2} (1+R^{-1})}\label{eq:A_p1q1}\\
e^{2B} \approx (1+8\pi \rho_0 r_0^{2}) - 8\pi \rho_0 r_0^{2} \frac{\ln{(R+1)}}{R}\label{eq:B_p1q1} 
\end{gather}
for some constant $C$.

The definition of the geodesic curvature $\eqref{eq:kappa}$ involves the covariant derivative operator $\nabla$. For the circular arc, it turns out we do not need all the connection coefficients $\Gamma_{jk}^i$ to solve the geodesic curvature. The unit tangent and unit acceleration vector of the circular arc $\gamma$ are 
\begin{eqnarray}
\dot{\gamma} &= \partial_{\phi} / \sqrt{g_{\phi\phi}} \label{eq:unit_tangent}\\
\ddot{\gamma} &= \partial_{r} / \sqrt{g_{rr}} \label{eq:unit_accel}
\end{eqnarray}
respectively. So the geodesic curvature is simply
\begin{equation}\label{eq:kappa_solve}
\kappa = g_{rr}(\nabla_{\dot{\gamma}}\dot{\gamma})^r\ddot{\gamma}^r
\end{equation}
where
\begin{equation}\label{eq:nabla_r}
(\nabla_{\dot{\gamma}})^r=\Gamma^{r}_{\phi \phi}\dot{\gamma}^{\phi}\dot{\gamma}^{\phi}
\end{equation}
and
\begin{equation}\label{eq:g_phiphi_r}
g_{\phi\phi, r} = \frac{2g_{\phi\phi}}{r} \Bigg(1 - 4\pi\rho_0 r_0^2\bigg(1-\frac{\ln{(R+1)}}{R} \bigg) \Bigg)
\end{equation}
With the additional factor $dt/d\phi=g_{\phi\phi}^{1/2}$, the integrand of the line integral in equation \eqref{eq:master} is
\begin{equation}\label{eq:pre_integrand_lineintegral_p1q1}
\kappa\frac{dt}{d\phi} = \frac{1}{r}\sqrt{\frac{g_{\phi\phi}}{g_{rr}}}\Bigg(1 - 4\pi\rho_0 r_0^2\bigg(1-\frac{\ln{(R+1)}}{R} \bigg) \Bigg)
\end{equation}
And taking the limit $r\rightarrow\infty$, we find
\begin{equation}\label{eq:integrand_lineintegral_p1q1}
\lim_{r\rightarrow\infty} \kappa\frac{dt}{d\phi} = 1 - 8\pi\rho_0 r_0^2
\end{equation}

Finally from equations \eqref{eq:area_part_p1q1} and \eqref{eq:integrand_lineintegral_p1q1}, the deflection angle is 
\begin{multline}\label{eq:deflect_p1q1}
\delta(B) = 8\pi\rho_0r_{0}^2\bigg(\pi-\frac{I_3}{B}\bigg) \\= \frac{16\pi\rho r_{0}^2}{B} \bigg(-\ln \frac{B}{2} + \sqrt{B^2-1}\arctan{\sqrt{B^2-1}}\bigg)
\end{multline}
Here, $\delta(0)=0$ and $\delta(\infty)=8\pi^2\rho_0r_{0}^2$. Note that the deflection at $B=\infty$ is non-vanishing. We will discuss this in Section \ref{sec:discussion}.

\subsection{Case $(0,\beta\geq4)$}
All the remaining cases are solved in the same manner as the previous ones, so we will only present results from here on.

For this case, we have
\begin{equation}\label{eq:rho_p0q}
    \rho(R)=\frac{\rho_0}{(R+1)^\beta}
\end{equation}
\begin{multline}\label{eq:mass_p0q}
    m(R) = 4 \pi \rho_{0} r_{0}^3 \Bigg(\bigg( \frac{1}{\beta-3}-\frac{2}{\beta-2}+\frac{1}{\beta-1}\bigg)\\-\frac{1}{\beta-3}\frac{1}{(R+1)^{\beta-3}}+\frac{2}{\beta-2}\frac{1}{(R+1)^{\beta-2}}\\-\frac{1}{\beta-1}\frac{1}{(R+1)^{\beta-1}} \Bigg)
\end{multline}
\begin{multline}\label{eq:K_p0q}
    \mathcal{K} =  8\pi \rho_{0} r_{0}^2 \Bigg(-\frac{R}{(R+1)^\beta}-\frac{1}{\beta-3}\frac{1}{(R+1)^{\beta-1}}\\+\frac{2}{(\beta-3)(\beta-2)(\beta-1)}\sum_{i=2}^{\beta-1}\frac{i-1}{(R+1)^i} \Bigg)
\end{multline}
\begin{multline}\label{eq:deflect_p0q}
    \delta(B) = \frac{8\pi \rho_{0} r_{0}^2}{(\beta-1)(\beta-2)(\beta-3)} \Bigg(\frac{4}{B} \\-2\frac{I_1^{\beta-1}}{B}+(\beta-4)(\beta-1)I_2^{\beta-1}\\-(\beta-2)(\beta-1)I_2^{\beta-2}\Bigg)
\end{multline}
One can check that $\delta(0)=\delta(\infty)=0$.

\subsection{Case $(0,3)$}
Here,
\begin{equation}\label{eq:rho_p0q3}
    \rho(R)=\frac{\rho_0}{(R+1)^3}
\end{equation}
\begin{multline}\label{eq:mass_p0q3}
    m(R) = 4\pi\rho_{0}r_{0}^3 \bigg(\ln{(R+1)}+\frac{2}{R+1}\\-\frac{1}{2}\frac{1}{(R+1)^2}-\frac{3}{2}\bigg)
\end{multline}
\begin{equation}\label{eq:deflect_p0q3}
    \delta(B) = 8\pi \rho_{0} r_{0}^2\bigg(\frac{1}{2}I_2^2 - \frac{3}{2}I_2^1 + \frac{I_3}{B}\bigg)
\end{equation}
Similarly, $\delta(0)=\delta(\infty)=0$.

\subsection{Case $(0,2)$}
This density is of the type $\beta=2$, so the calculation is similar to the case $(1,1)$. With the density and mass functions
\begin{equation}\label{eq:rho_p0q2}
    \rho(R)=\frac{\rho_0}{(R+1)^2}
\end{equation}
\begin{multline}\label{eq:mass_p0q2}
    m(R) = 4\pi\rho_{0}r_{0}^3 \bigg(-2\ln{(R+1)}-\frac{1}{R+1}\\-\frac{1}{2}\frac{1}{(R+1)^2}+R+1\bigg)
\end{multline}
the area integral in equation \eqref{eq:master} gives
\begin{equation}\label{eq:area_part_p0q2}
    \delta_{\textrm{area}}(B) = 8\pi \rho_{0} r_{0}^2\bigg(I_2^1-2\frac{I_3}{B}\bigg)
\end{equation}

For the line integral part, we calculate first the metric coefficients:
\begin{equation}\label{eq:A_p0q2}
e^{2A} \approx C^{-2} (R+1)^{8\pi \rho_0 r_0^{2} (1-2R^{-1})}
\end{equation}
\vspace{-20pt}
\begin{multline}\label{eq:B_p0q2}
e^{2B} \approx (1+8\pi \rho_0 r_0^{2})-8\pi \rho_0 r_0^{2}\bigg(\frac{1}{R+1}-\frac{2\ln{(R+1)}}{R}\bigg)
\end{multline}
Solving equation \eqref{eq:kappa}, we get
\begin{equation}\label{eq:pre_integrand_lineintegral_p0q2}
\kappa\frac{dt}{d\phi} = \frac{1}{r}\sqrt{\frac{g_{\phi\phi}}{g_{rr}}}\Bigg(1 - 4\pi\rho_0 r_0^2\bigg(\frac{R-2}{R+1}+\frac{2\ln{(R+1)}}{R} \bigg) \Bigg)
\end{equation}
Coincidentally, equation \eqref{eq:pre_integrand_lineintegral_p0q2} follows the same limit in equation \eqref{eq:integrand_lineintegral_p1q1}. Finally, we get the deflection angle from equations \eqref{eq:area_part_p0q2} and \eqref{eq:integrand_lineintegral_p1q1}:
\begin{equation}\label{eq:deflect_p0q2}
    \delta(B) = 8\pi\rho_0r_{0}^2\bigg(\pi-2\frac{I_3}{B}+I_2^1\bigg)
\end{equation}
Just like the previous case $(1,1)$, we have $\delta(0)=0$ and $\delta(\infty)=8\pi^2\rho_0r_{0}^2$.

\section{Discussion}\label{sec:discussion}
The deflection angles computed from the energy density sequence $(\alpha,\beta)$ are well characterized by their $b\rightarrow 0$ and $b \rightarrow \infty$ limits. These limiting behaviors reflect the geometry of the optical manifold at the center and the asymptotic regions, respectively. The only two limiting behaviors are (1) a vanishing deflection angle corresponding to a locally flat region, and (2) a non-zero deflection angle limit determined by a conical structure at the region. Table \ref{tab:summary} gives a summary of the deflection angle limits.
\begin{widetext}
\vspace{-20pt}
\begin{center}
\begin{table}[h!]
    \caption{Summary of deflection angle limits at zero and infinite impact parameter}
    \begin{tabularx}{0.75\textwidth}{zzzzz}\hline\hline
        &$\beta=2$&$\beta=3$&$\beta\geq4$\\
        &$m_{\infty}$ undefined, asymptotically conical,&$m_{\infty}$ undefined, asymptotically flat,&$m_{\infty}$ defined, asymptotically flat,\\
        &$\delta(\infty)=8\pi^2\rho_0r_{0}^2$&$\delta(\infty)=0$&$\delta(\infty)=0$\\\midrule
        $\alpha=0$, flat center, $\delta(0)=0$&&&Dehnen-type $(\beta=4)$, Plummer-like $(\beta=5)$\\\midrule
        $\alpha=1$, flat center, $\delta(0)=0$&&NFW&Hernquist $(\beta=4)$\\\midrule
        $\alpha=2$, conical center, $\delta(0)=8\pi^2\rho_0r_{0}^2$&singular isothermal sphere&&Jaffe $(\beta=4)$\\\hline
    \end{tabularx}
    \label{tab:summary}
\end{table}
\end{center}
\vspace{-20pt}
\end{widetext}

Vanishing deflection at zero impact parameter in an asymptotically flat spacetime is guaranteed when there are no singularities anywhere. By the angular symmetry of the metric, the geodesic light trajectory at $b=0$ must be the straight curve in flat space: $(x\circ\Gamma)(\lambda(t))=x_\textrm{S}-(\lambda(t)/\lambda_\textrm{O})(x_\textrm{S}-x_\textrm{O})$ in Cartesian coordinates, where $t$ is the same arclength parameter in equation \eqref{eq:metric_opt}. Thus, the region $D$ in the central equation \eqref{eq:master} is just the upper half-disk centered at $C$ where $S$ and $O$ are at the vertices. Taking the limit $x_\textrm{S},\,x_\textrm{O}\rightarrow\infty$, it is no surprise from the left-hand side of equation \eqref{eq:master} that $\delta(0)=0$. This may seem trivial at first, but it is instructive to present this argument because 1) it will not be apparent from the Gaussian curvature integral in equation \eqref{eq:deflect_area} that it will vanish at $b=0$, and 2) one cannot apply the same reasoning when there is a singularity at $C$.

The central geometry of densities with $\alpha=2$ approaches that of the singular isothermal sphere (SIS). The singular center prohibits the direct application of the Gauss-Bonnet theorem when the impact parameter is exactly zero. That is, one can only use a region $D$ that approaches the upper-half disk, but one cannot use the latter directly. In fact, the Gaussian curvature integral approaches a non-zero value as $b\rightarrow0$. This value is related to the central geometry of the optical manifold. We see here how the Gauss-Bonnet method highlights the role of topology in the contrasting central behavior of $\alpha=0,1$ and $\alpha=2$ densities.

Gibbons and Werner \cite{Gibbons2008} showed that the embedding of the low-density SIS optical manifold, with isotropic velocity dispersion $\sigma^2$, in flat $\mathbb{R}^3$ charted by $(s,\phi,z)$ is the cone
\begin{equation}\label{eq:sis_cone}
    z = \sqrt{8\sigma^2}\,\frac{\big(1-\frac{9\sigma^2}{2}\big)^{1/2}}{1-6\sigma^2}s
\end{equation}
In terms of $\rho_0$ and $r_0$, $\sigma^2 = 2 \pi \rho_{0} r_{0}^2$. A cone $z=ks$ has a deficit angle $\Delta$ of
\begin{equation}\label{eq:deficit_exact}
\Delta = 2\pi \bigg( 1 - \sqrt{\frac{1}{1+k^2}} \bigg)
\end{equation}
(derived in Section \ref{sec:appendix}). Up to first order in $\mu$, the SIS deflection angle is
\begin{equation}
    \delta_{\textrm{SIS}}=\Delta/2=8\pi^2\rho_0 r_0^2
\end{equation}
That is, the constant SIS deflection angle is half the deficit angle of its conical optical manifold. From this point of view, deflection by SIS is entirely topological, i.e., due to the deficit angle of the conical manifold. We notice that this is the same value of the zero impact parameter limit of deflection by $\alpha=2$ densities. This suggests that such deflections are also due to the conical center of the $\alpha=2$ optical manifolds.

The infinite impact parameter behavior of deflection is more apparent to see. For asymptotically flat spacetimes, it is clear from equation \eqref{eq:deflect_area} that the integral must vanish as the lower bound of $R$ approaches the upper bound. Meanwhile, $\beta=2$ densities approach SIS distribution at large radial distances, thus the embedding in flat $\mathbb{R}^3$ of this region of the optical manifold is also approximately conical. As expected, we get a deflection angle of $\delta(\infty)=\Delta/2=8\pi^2\rho_0 r_0^2$, similar to the case of $\alpha=2$ when $b=0$.

\begin{figure}[h!]
    \centering
    \begin{minipage}{0.45\textwidth}
    \includegraphics[width=\textwidth]{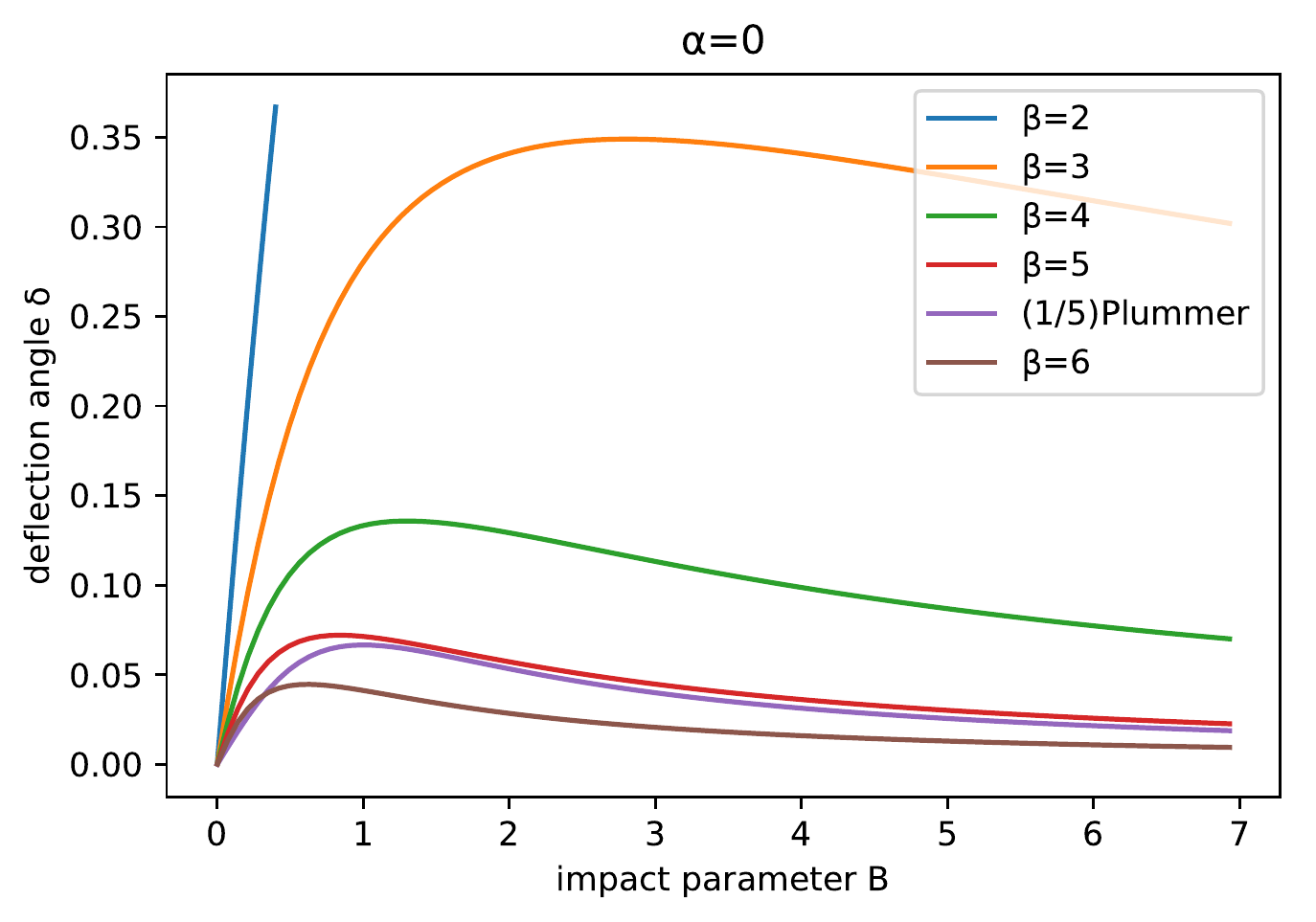}
    \end{minipage}
    \begin{minipage}{0.45\textwidth}
    \includegraphics[width=\textwidth]{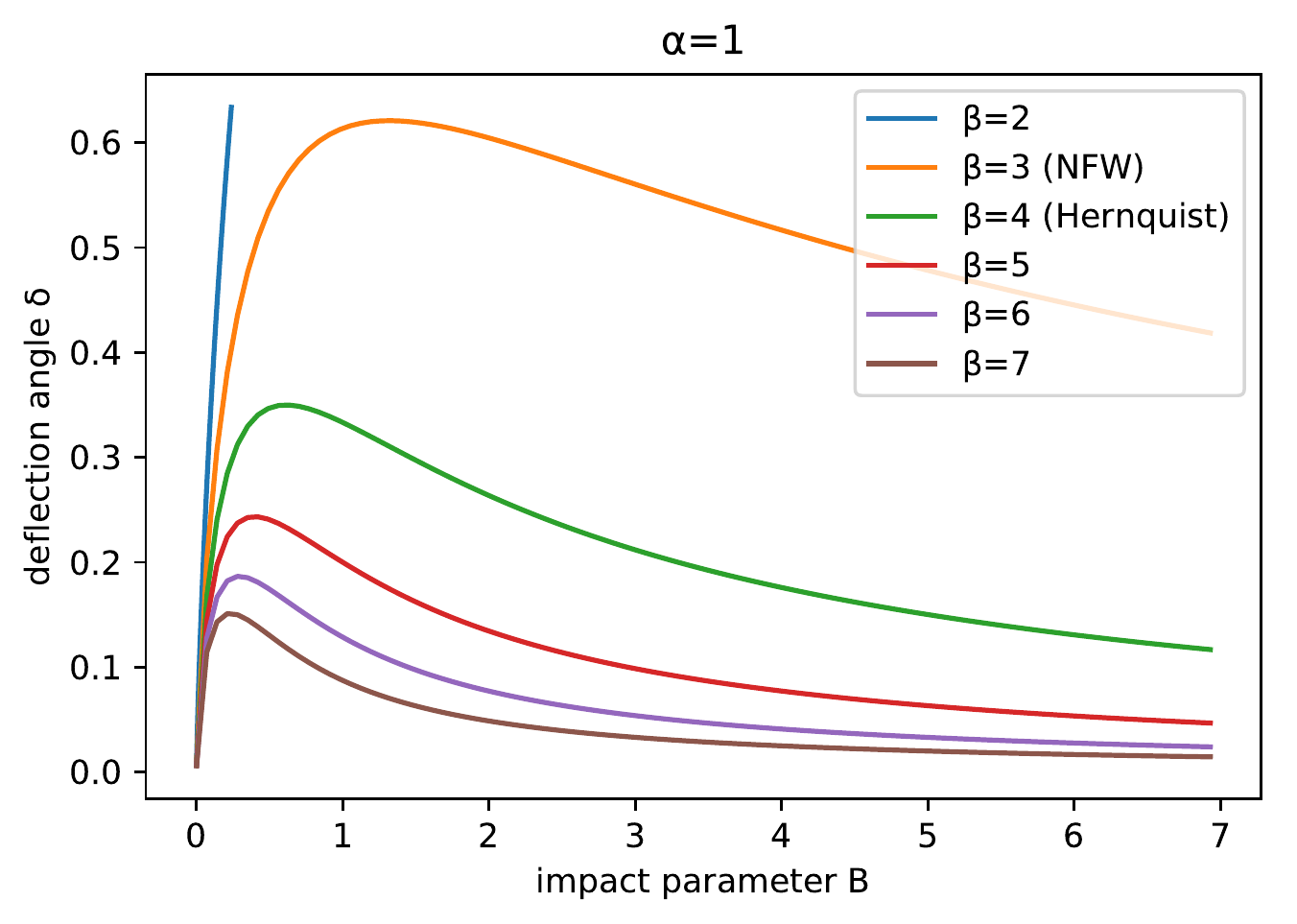}
    \end{minipage}
    \begin{minipage}{0.45\textwidth}
    \includegraphics[width=\textwidth]{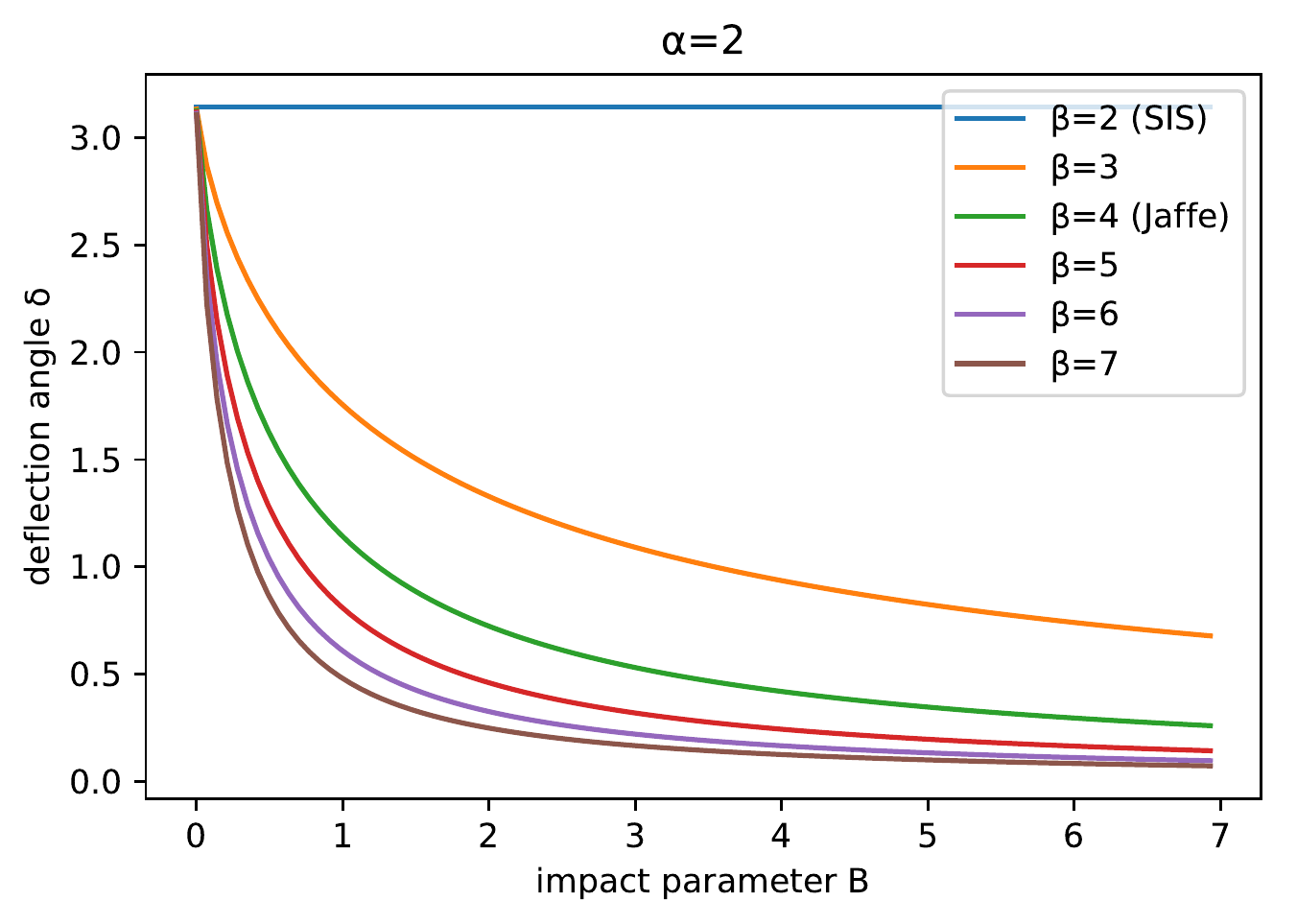}
    \end{minipage}
    \caption{Plots of deflection angles in factors of $8\pi\rho_0 r_0^2$ as a function of the scaled impact parameter $B=b/r_0$. The Plummer curve is scaled down by $1/5$ to emphasize its resemblance to the $(0,5)$ curve.}
    \label{fig:plots}
\end{figure}

Plots of deflection angles are shown in Figure \ref{fig:plots}. We see that deflection by $\beta=2$ densities always approach the angle $\Delta/2$ as the impact parameter grows large. Meanwhile, $\beta=1$ deflection falls off considerably slower than $\beta\geq4$ due to logarithmic terms plaguing the decay. We also note that since the density function of $(0,5,1)$ behaves comparable to the Plummer sphere $(0,5,1/2)$, the Plummer deflection (e.g., \cite{Gibbons2008}) follows roughly the deflection curve of $(0,5,1)$ given a suitable scale factor.

\section{Conclusions and Recommendations}
To summarize, we have obtained new analytic expressions for the first-order deflection angle due to spherical two-power law densities $(\alpha,\beta,1)$ in equation \eqref{eq:rho_general} for $\alpha=0,1,2$ and $\beta=2,3,4,\ldots$ applicable for low-density distributions $\mu=\rho_0r_0^2\ll1$ using the method by Gibbons and Werner \cite{Gibbons2008}. Our main results are presented in equation \eqref{eq:deflect_p2q} for the case $(2,\beta\geq2)$, \eqref{eq:deflect_p2q1} for $(2,1)$, \eqref{eq:deflect_p1q} for $(1,\beta\geq3)$, \eqref{eq:deflect_NFW} for $(1,2)$, \eqref{eq:deflect_p1q1} for $(1,1)$, \eqref{eq:deflect_p0q} for $(0,\beta\geq4)$, \eqref{eq:deflect_p0q3} for $(0,3)$, and \eqref{eq:deflect_p0q2} for $(0,2)$. Explicit forms are determined for the named densities: Hernquist \eqref{eq:deflect_Hern}, NFW, and Jaffe \eqref{eq:deflect_Jaffe}. Our calculated Hernquist and NFW deflections are consistent with the result of Keeton \cite{keeton2001catalog} and Bartelmann \cite{bartelmann1996arcs}, respectively. Our calculations demonstrate how the Gauss-Bonnet method can be more convenient for non-relativistic and spherically symmetric distributions compared to canonical methods, such as integration from $g(p^{\mu},p^{\mu})$ and the method of the thin lens approximation.

We have demonstrated that the Gibbons-Werner insight into the role played by topology in gravitational lensing extends to mass distributions beyond those that the authors initially considered, as is explicitly demonstrated in the $\alpha=0,1$ and $\alpha=2$ densities.  We have shown how the topological properties of the corresponding optical manifold immediately imply vanishing deflection of the former cases and finite deflection for the latter case in the limit of vanishing impact parameter ($b\rightarrow0$). The topology-controlled behavior of the deflection also obtains in the $b\rightarrow \infty$ limit. Densities with $\alpha=0,1$ and $\beta>2$ have vanishing deflection at $b=0$ and $b=\infty$, respectively. This is due to the centrally flat optical manifold of the former, and the asymptotically flat manifold of the latter. Meanwhile, deflection of densities with $\alpha=2$ and $\beta=2$ approaches a finite value at $b=0$ and $b=\infty$, respectively. In these regions, these densities are well-approximated by the singular isothermal sphere distribution. Gibbons and Werner \cite{Gibbons2008} previously noted that the constant deflection angle of the low-density singular isothermal sphere takes the value of half the deficit angle of its conical optical manifold, suggesting that the deflection is due to the conical angle defect. Here, we find the same to be true for the limiting cases of $\alpha=2$ and $\beta=2$. Outside the limiting behaviors of the deflection, however, we emphasize that geometrical details of the optical manifold do play the dominant role.

An immediate extension of this study is to find a general expression that includes non-integer values of $\alpha$ and $\beta$, which may provide a better fit with certain galactic densities. Such values complicate the form of the integrals and render invalid the analytic techniques used here. A broader question of continuing interest is the possibility of isolating topological contributions from metric contributions to the deflection angle. We seek to address these and related questions in future work.

\section*{Acknowledgements}
K.D.L. is grateful to Gary Gibbons for helpful conversations that shaped the writing of this paper. This research is supported by the University of the Philippines OVPAA through Grant No.~OVPAA-BPhD-2016-13. 

\appendix\label{sec:appendix}
\section{Partial fraction decomposition: denominator with two distinct roots}

We wish to find the coefficients $A_i$ and $B_i$ that satisfies
\begin{equation}\label{eq:partialfracdec}
\frac{1}{(x-a)^p(x-b)^q} = \sum_{i=1}^{p} \frac{A_i}{(x-a)^i} + \sum_{j=1}^{q} \frac{B_j}{(x-b)^j}
\end{equation}
where $p$ and $q$ are positive integers. The partial fraction decomposition looks somehow similar to Laurent expansions. This suggests that the coefficients might be extracted from relevant Laurent series expansions. 
We start by appealing to this well known Kronecker delta expression as an isolation tool:
\begin{equation}\label{eq:isolate}
\frac{1}{2\pi i} \oint_{C_a} \frac{dz}{(z-a)^n} = \delta_{1n}
\end{equation}
where $C_a$ is a closed contour enclosing $z=a$ as the only singular point. Suppose we want to find the coefficient $A_k$. To utilize equation \eqref{eq:isolate}, we multiply both sides of equation \eqref{eq:partialfracdec} with $(x-a)^{k-1}$, so that 
\begin{multline}\label{eq:step1}
\frac{1}{(x-a)^{p-k+1}(x-b)^q} =\\ \sum_{i=-k+2}^{p-k+1} \frac{A_{i+k-1}}{(x-a)^i} + (x-a)^{k+1} \sum_{j=1}^{q} \frac{B_j}{(x-b)^j}
\end{multline}

\noindent Integrating both sides of equation \eqref{eq:step1} on the complex plane along the closed contour $C_a$, we find that on the right-hand side only the $i=1$ term survives on the first summation, as per relation \eqref{eq:isolate}, while the third summation vanishes because it is an analytic function on the domain enclosed by the contour. Thus, 
\begin{equation}\label{eq:basic}
A_{k} = \frac{1}{2\pi i}\oint_{C_{a}} \frac{1}{(x-a)^{p-k+1}(x-b)^q} \, dx
\end{equation}
This is easily solved by calculus of residues:
\begin{equation}\label{eq:res}
A_{k}=\frac{1}{(p-k)!}\Big( \frac{\partial}{\partial a}\Big)^{p-k}_{x=a}\,\frac{1}{(x-b)^q}
\end{equation}
\begin{equation}
A_{k}={{p+q-1-k}\choose{q-1}}\frac{(-1)^{p-k}}{(a-b)^{p+q-k}}
\end{equation}
The coefficients $B_j$ are obtained in the same manner:
\begin{equation}
B_{k} = {{p+q-1-k}\choose{p-1}} \frac{(-1)^{q-k}}{(b-a)^{p+q-k}}
\end{equation}

\section{Evaluating the $I_{\MakeLowercase{\textit{i}}}$ integrals}
Only the integrals $I_1$ and $I_3$ are sketched here. $I_2$ is solved in the same manner as $I_1$ with some slight modifications.
\subsection{$I_1$ and $I_2$}
We use complex integration to evaluate the integral $I_1$. First, note that
\begin{equation}\label{eq:complexintegration}
I_1^q = \int_{0}^{\pi} \frac{\sin{\phi}\,\,d\phi}{(1+B\csc{\phi})^{q}} = \Im{\Bigg[ \int_{0}^{\pi} \frac{e^{i\phi}\,\,d\phi}{(1+B\csc{\phi})^q}\Bigg]} =: \Im{[\mathcal{I}]}
\end{equation}
We instead deal with the integral $\mathcal{I}$. However, the exponent of the denominator still complicates the evaluation. As a way out, we proceed as follows:
\begin{align}
\mathcal{I} &= \frac{1}{B^{q}} \int_{0}^{\pi} \frac{e^{i\phi}\,\,d\phi}{(B^{-1}+\csc{\phi})^{q}}\nonumber\\
&= \frac{1}{B^{q}} \bigg( \int_{0}^{\pi} \frac{e^{i\phi}\,\,d\phi}{(a+\csc{\phi})^q}\bigg) _{a=1/B}\nonumber\\
&= \frac{1}{B^q} \frac{(-1)^{q-1}}{(q-1)!} \Big( \frac{\partial}{\partial a}  \Big) _{a=1/B}^{q-1} \bigg(\int_{0}^{\pi} \frac{e^{i\phi}\,\,d\phi}{a+\csc{\phi}}\bigg)\label{eq:ez}
\end{align}
Let the integral in equation \eqref{eq:ez} be $\mathcal{J}$. The overall form of $\mathcal{J}$ suggests that the contour of the integral in the complex plane is a semi-circular arc centered at $z=0$ of unit modulus. Hence, we consider the integral
\begin{equation}\label{eq:K}
\mathcal{K} = \frac{1}{i} \oint_{\mathcal{C}} \frac{z^2-1}{az^2+2iz-a} \, dz
\end{equation}
where $\mathcal{C}=\mathcal{C}_1+\mathcal{C}_2$ is the contour traversed in the positive sense given by
\begin{eqnarray}\label{eq:c1}
\mathcal{C}_1: z(x) = x, x \in [-1,1] \\
\mathcal{C}_2: z(\phi) = e^{i\phi}, \phi \in [0,\pi]
\end{eqnarray}
The simple poles are at
\begin{equation}
z_{\pm} = i\bigg( -\frac{1}{a} \pm \sqrt{\frac{1}{a^2}-1}\bigg)
\end{equation}
Notice that the poles are never inside the domain enclosed by the contour $\mathcal{C}$, so $\mathcal{K}=0$, and
\begin{equation}
\mathcal{K} = \oint_{\mathcal{C}_1} (\ldots) + \oint_{\mathcal{C}_2} (\ldots) = 0
\end{equation}
The $\mathcal{C}_2$-integral is already $\mathcal{J}$, so
\begin{equation}\label{eq:last_int}
\mathcal{J} = -\frac{1}{i} \int_{-1}^{1}  \frac{x^2-1}{ax^2+2ix-a} \, dx
\end{equation}

We consider first the case $0<a<1$ where the poles are purely imaginary. Later on, we will argue that the answer we get here is the same as when $a>1$ by invoking the uniqueness of analytic continuation of functions; we expect $\mathcal{J}$ to be a well-behaved function of $a$. Evaluating $\mathcal{J}$ will finally solve $I_1^q$ (initial calculation will give the answer in terms of inverse tangent functions).

\subsection{$I_3$}
Here, we evaluate the interal $I_3$
\begin{equation}\label{eq:integral_j}
    I_3 = \int_{0}^{\pi} \sin{\phi}\ln{(1+B\csc{\phi})} \, d\phi
\end{equation}
Integrating by parts once, we proceed as
\begin{align}
    I_3 &= (-\cos{\phi}\ln{(1+B\csc\phi)})\Big\rvert_{0}^{\pi} - \int_{0}^{\pi} \frac{B\cot^2\phi}{B\csc\phi+1} \, d\phi \nonumber \\
                &= (\ldots)\Big\rvert_{0}^{\pi} - \frac{1}{B}\int_{0}^{\pi} \frac{B^2(\csc^2\phi - 1)}{1+B\csc{\phi}} \, d\phi \nonumber \\
                &= (\ldots)\Big\rvert_{0}^{\pi} - \frac{1}{B}\int_{0}^{\pi} \frac{(B^2\csc^2\phi - 1) + (1 - B^2)}{1+B\csc{\phi}} \, d\phi \nonumber \\
                &= (\ldots)\Big\rvert_{0}^{\pi} -\frac{1}{B}\int_{0}^{\pi} (B\csc\phi-1) \, d\phi - \bigg(\frac{1-B^2}{B}\bigg)I_2^1 \nonumber \\
                &= (\ldots)\Big\rvert_{0}^{\pi} + \ln{(\csc\phi+\cot\phi)}\Big\rvert_{0}^{\pi} + \frac{\pi}{B} - \bigg(\frac{1-B^2}{B}\bigg)I_2^1 \nonumber \\
                &= \ln{\bigg(\frac{\csc\phi+\cot\phi}{(B\csc\phi+1)^{\cos\phi}}\bigg)}\Bigg\rvert_{0}^{\pi} + \frac{\pi}{B} - \bigg(\frac{1-B^2}{B}\bigg)I_2^1 \nonumber \\
                &=\pi B + 2\ln \frac{B}{2} - 2\sqrt{B^2-1}\arctan{\sqrt{B^2-1}}
\end{align}
It should be noted that the first two terms on the sixth line are undefined individually; only their sum has a finite limit at $\phi=0$ and $\phi=\pi$.

\section{Deficit angle of a cone}
The deficit angle of a cone is defined by its slope $k$. Consider the cone $z-ks=0$ in flat $\mathbb{R}$ charted by cylindrical coordinates $(s,\phi,z)$. The induced metric on the cone is
\begin{align}
    ds^2_{\textrm{cone}} &= ds^2 + s^2d\phi^2 + d(ks)^2\\
    &= (k^2+1)(ds^2 + s^2d\Tilde{\phi}^2)
\end{align}
where
\begin{equation}
    \Tilde{\phi} = \frac{\phi}{\sqrt{k^2+1}}
\end{equation}
We see that the metric on the conical surface is conformal to the Euclidean metric, but with a reduced angular range: $\Tilde{\phi} \in [0,2\pi/\sqrt{k^2+1})$. Thus, the deficit angle is
\begin{equation}\label{eq:deficit_exact}
\Delta = 2\pi \bigg( 1 - \sqrt{\frac{1}{k^2+1}} \bigg) = \pi k^2 + \mathcal{O}(k^4)
\end{equation}

\bibliography{references.bib}
\end{document}